\documentclass[twoside,slac_one]{revtex4}
\usepackage{graphicx}
\usepackage{fancyhdr}
\usepackage{amsmath} % American Mathematics Society standards
\usepackage{bm}% bold math
\usepackage{amsxtra}
\usepackage{amssymb}
\usepackage{amsthm}
\usepackage{latexsym}
\usepackage{lscape}

\begin{document}

\title{Probing Scalar Mesons Below and Above 1 GeV}


\author{Amir H. Fariborz}
\affiliation{Department of Engineering, Science and Mathematics, State University of New York Institute of Technology, Utica, NY 13502, USA}

\begin{abstract}
Within the context of a generalized linear sigma model that includes two nonets of scalar mesons (a two-quark nonet and a four-quark nonet) and two nonets of pseudoscalar mesons (a two-quark nonet and a four-quark nonet) a collective description of scalar and pseudoscalar mesons below and above 1 GeV  is studied.  The quark contents of these states are probed; estimates on their quark components are extracted; and prediction for $\pi \pi$ scattering amplitude is made.  An overview of these studies is presented here.
\end{abstract}

\maketitle

\thispagestyle{fancy}


%%%%%%%%%%%%%%%%%%%%%%%%%%%%%%%%%%
\section{Introduction}

Scalar mesons are important states in hadron spectroscopy \cite{pdg} because they are related to the Higgs bosons of QCD,  induced spontaneous chiral symmetry breaking and probe QCD vacuum.   They are also important intermediate states in several low-energy processes such as $\pi\pi$, $\pi K$, $\pi \eta$ scatterings as well as several decays such as $\eta$ and $\eta'$ decays, as well as many heavier meson decays such as, for example,  the semileptonic decays of $D_s$ meson.   However, understanding their properties is known to be nontrivial:
Experimentally,  some of these states are very broad and therefore interfere with the nearby states, and theoretically,  they do not fit into the conventional SU(3) multiplets,  something that is know to work fairly well for other light hadrons (such as the vector mesons).  As a result scalars mesons have been the topic of intense investigation in low-energy QCD \cite{j77,general_refs}.

\begin{figure}[ht]
\centering
\includegraphics[width=120mm]{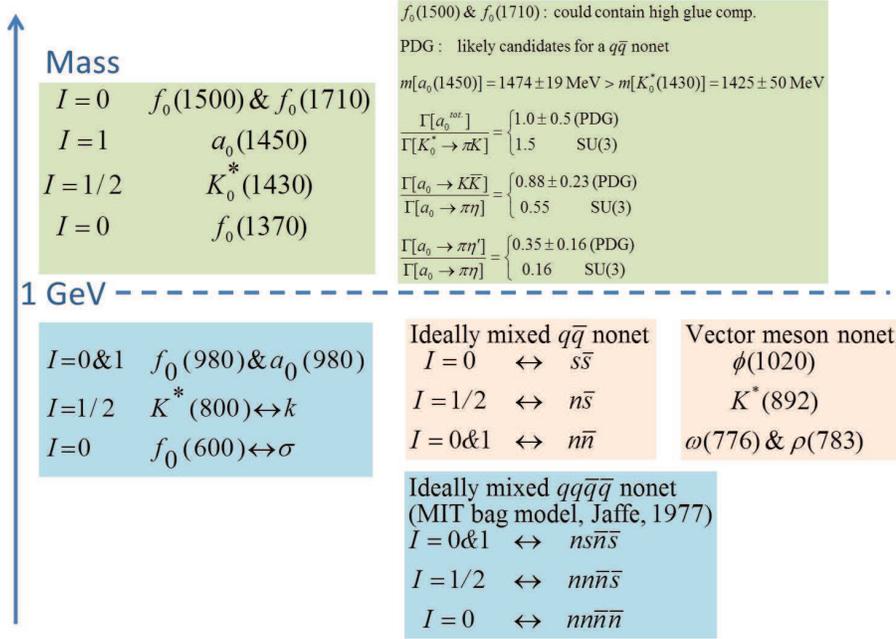}
\caption{The two boxes on the left list the scalar mesons below and above 1 GeV.  The box in the middle shows the isospin and the quark substructure of an ideally-mixed quark-antiquark nonet (which is the reverse of the scalar meson spectrum below 1 GeV, but quite consistent with the vector meson nonet).   The lowest box shows the isospin and the quark substructure of an ideally-mixed two quark-two antiquark nonet (which clearly has the same structure to that of scalar states below 1 GeV).   The top right box, provides some of the properties of the scalar mesons above 1 GeV which are not quite consistent with the assumption that these states are pure quark-antiquark objects. }
\label{Fscalarstates}
\end{figure}

Below 1 GeV the known states are listed in Fig. \ref{Fscalarstates}:    the light and broad isosinglet $f_0(600)$ or sigma, followed by isobublet $K^*(800)$ or kappa meson and the two nearly degenerate  states, isosinglet $f_0(980)$ and isotriplet $a_0(980)$.
One can  immediately see that these states do not quite follow a quark-antiquark spectroscopy:
First, if they were quark-antiquark states, one would expect their masses to be close to the axial vector meson masses around 1.2 GeV, but clearly these are all below 1 GeV.
Second,   if we attempt to collect them into a quark-antiquark nonet, we find that it does not quite work.   In fact the mass ordering in a pure ideally mixed $q {\bar q}$ nonet is completely the opposite of that of scalars (see Fig. \ref{Fscalarstates}).   Such ideally mixed $q {\bar q}$  nonet is known to work for vector mesons that have the same natural mass ordering.
As a fundamental solution to these problems,  the four-quark model (i.e two-quark two-antiquark) for these states was proposed in MIT bag model \cite{j77}.    Fig. \ref{Fscalarstates} shows the mass ordering for an ideally mixed four-quark nonet which agrees with the ordering of the light scalars and therefore seems to be a natural template for these states.  However,  the ideally mixed four-quark picture, even though provides a consistent picture for the mass spectrum, but has deviations from some of the experimental decay properties of scalar mesons.

Above 1 GeV,  the known scalar states are also listed in Fig. \ref{Fscalarstates}:   The isosiglet state $f_0(1370)$ followed by the isodoublet $K_0^*(1430)$, the isotriplet $a_0(1450)$ and the two isosinglet states $f_0(1500)$ and $f_0(1710)$.    The $f_0(1370)$ has a large uncertainty on its mass and decay width \cite{pdg} and it could be anywhere in the 1.2 to 1.5 GeV range.   The two isosinglet states $f_0(1500)$ and $f_0(1710)$ are speculated to contain a considerable glue component.  The scalar states above 1 GeV are generally believed to form a quark-antiquark nonet, even thought there are some deviations from this picture: For example, if $K_0^*(1430)$ and $a_0(1450)$ belong to the same $q{\bar q}$ nonet, then why should $a_0$ (that should not have a strange quark) be heavier than $K_0^*$ which has one strange quark?   Also their decay ratios do not quite follow SU(3) patterns:   In Fig. \ref{Fscalarstates} the SU(3) predictions for various decay ratios are compared with the experimental estimates and we see that even though the  order of magnitudes are consistent but there are some deviations.

To recap, the scalar states below 1 GeV seem to be close to four-quark states (with some deviations) and those above 1 GeV seem be close to two-quark states (with some deviations).   To generate such deviations from pure four-quark and pure two-quark pictures, it seems natural to investigate an underlying mixing among four- and two-quark states.   This is the main objective of the present discussion.     In the work of ref. \cite{Mec} such a mixing was studied within a nonlinear chiral Lagrangian framework and it was shown that it can provide a consistent picture for the properties of scalar mesons below and above 1 GeV. The same framework was further extended to the case of isosinglet scalar states in \cite{Mec_f} and coherent results for these states were obtained.  Similar mixing patterns have also been studied by other investigators \cite{mixing}.

\begin{figure*}[ht]
\centering
\includegraphics[width=100mm]{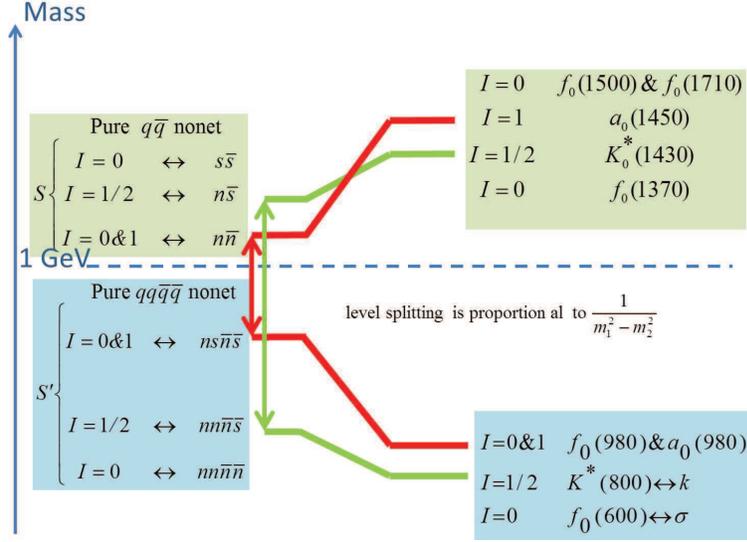}
\caption{Mixing mechanism between a two-quark scalar meson nonet $S$ and a four-quark scalar meson nonet $S'$.} \label{F_mass_splitting}
\end{figure*}

The same types of mixing, similar to those of  refs. \cite{Mec, Mec_f}, has been also investigated within the context of linear sigma model in refs. \cite{ulsm,toy_model,fjs_glsm,global} which will be reviewed in some details in the present work.  We see in Fig. \ref{F_mass_splitting} the general idea of this mixing mechanism for the scalar sector:   Starting out with two ``bare'' (unmixed) scalar meson nonets, a pure four-quark nonet below 1 GeV ($S'$)  and a pure two-quark nonet above 1 GeV ($S$) with the internal isospins and mass spectrum given for each nonet in the figure.    This mixing mechanism  was first applied in \cite{Mec} to explain the mass spectrum and decay properties of the $a_0(1450)$ and $K_0^*(1430)$ mentioned above.    Allowing the two nonets to mix with each other, the  properties of $a_0(1450)$ and $K_0^*(1430)$ can be naturally explained in the following way.      According to the general mixing property of two states with bare (unmixed) masses $m_1$ and $m_2$, it can be easily seen that the mixing leads to splitting of the physical masses away from the ``bare'' masses and this splitting is inversely proportional to the ``bare'' mass difference. As  Fig.
\ref{F_mass_splitting} shows,  the two bare isotriplet states are closer to each other (compared to the two isodoublet states), and as a result, when we allow these states to mix, the isotriplet states split more than the isodoublets and  therefore there is a level crossing that naturally explains why, for example,  $a_0(1450)$ is heavier than the $K_0^*(1430)$.  Moreover, some of the decay properties of these states can be naturally understood based on such a mixing scenario \cite{Mec}.

In this article, a brief review of this mixing mechanism within the context of a generalized linear sigme model is presented. We give  a brief review of the Lagrangian in the next section, followed by a summary of the numerical results in Sec. III and a short summary and discussion in Sec. IV.
%%%%%%%%%%%%%%%%%%%%%%%%%%%%%%%%%%
\section{The Lagrangian}
The two scalar nonets $S$ and $S'$ are combined with their pseudoscalar partners to form the chiral nonets $M$ and $M'$:
\begin{eqnarray}
M&=& S + i \phi,\\
M'&=& S' + i \phi',
\end{eqnarray}
that transform under SU(3)$_{\rm L} \times$ SU(3)$_{\rm R} \times$ U(1)$_{\rm A}$ as:
\begin{eqnarray}
M&\rightarrow& e^{2i\nu}\, U_{\rm L} M U^\dagger_{\rm R},\\
M'&\rightarrow& e^{-4i\nu}\, U_{\rm L} M' U^\dagger_{\rm R}.
\end{eqnarray}
These transformations can be easily verified in terms of the quark fields inside $M$ and $M'$. The quark-antiquark substructure of $M$ can be written as:
\begin{equation}
M_a^b = {\left( q_{bA} \right)}^\dagger \gamma_4 \frac{1 + \gamma_5}{2} q_
{aA},
\label{M}
\end{equation}
where $a$ and $A$ are respectively flavor and color indices.  For $M'$ there are three possibilities:  First, to write $M'$ as a molecule of two $M$'s, i.e.
\begin{equation}
M_a^{(2)b} = \epsilon_{acd} \epsilon^{bef}
 {\left( M^{\dagger} \right)}_e^c {\left( M^{\dagger} \right)}_f^d.
\label{M2}
\end{equation}
The second and the third substructures for $M'$ correspond to two different ways that  two quarks and two antiquarks can be combined to form a nonet.   Depending on whether the two quarks and the two antiquarks have spin 0 or 1, we have two possibilities.   For spin 0:
\begin{equation}
M_g^{(3)f} = {\left( L^{gA}\right)}^\dagger R^{fA},
\end{equation}
with
\begin{eqnarray}
L^{gE} = \epsilon^{gab} \epsilon^{EAB}q_{aA}^T C^{-1} \frac{1 +
 \gamma_5}{2} q_{bB}, \nonumber \\
R^{gE} = \epsilon^{gab} \epsilon^{EAB}q_{aA}^T C^{-1} \frac{1 - \gamma_5}{2}
 q_{bB}.
\end{eqnarray}
For spin 1:
\begin{equation}
M_g^{(4) f} = {\left( L^{g}_{\mu \nu,AB}\right)}^\dagger R^{f}_{\mu \nu,AB},
\end{equation}
with
\begin{eqnarray}
L_{\mu \nu,AB}^g = L_{\mu \nu,BA}^g = \epsilon^{gab} q^T_{aA} C^{-1}
 \sigma_{\mu \nu} \frac{1 + \gamma_5}{2} q_{bB}, \nonumber \\
R_{\mu \nu,AB}^g = R_{\mu \nu,BA}^g = \epsilon^{gab} q^T_{aA} C^{-1}
 \sigma_{\mu \nu} \frac{1 - \gamma_5}{2} q_{bB}.
\end{eqnarray}
(See refs. \cite{fjs_glsm} for a detailed description.)   It can be shown that out of the three quark substructures for $M'$, only two are independent.

The general structure of the Lagrangian is:
\begin{equation}
{\cal L} = - \frac{1}{2} {\rm Tr}
\left( \partial_\mu M \partial_\mu M^\dagger
\right) - \frac{1}{2} {\rm Tr}
\left( \partial_\mu M^\prime \partial_\mu M^{\prime \dagger} \right)
- V_0 \left( M, M^\prime \right) - V_{SB},
\label{mixingLsMLag}
\end{equation}
where $V_0(M,M^\prime) $ stands for a function made
from SU(3)$_{\rm L} \times$ SU(3)$_{\rm R}$
(but not necessarily U(1)$_{\rm A}$) invariants
formed out of
$M$ and $M^\prime$ and $V_{SB}$ is the flavor symmetry breaking term.
In dealing with this Lagrangian,  two approaches have been considered:

Approach 1 is based on the underlying chiral symmetry and the resulting generating equations \cite{toy_model}. In this approach no specific choice for the chiral invariant part of $V_0$ is made,  and only the axial anomaly, the $V_{SB}$ and the condensates are modeled.   The details of this approach can be found in ref. \cite{toy_model} and are not discussed here.    The main conclusions are:
(i) The light pseudoscalars, as expected from conventional phenomenology, remain dominantly close to quark-antiquark states;  (ii)  The kappa meson tends to become a dominantly four-quark state; and (iii) This approach  does not provide any information on $a_0$ and $f_0$ systems.   To study  $a_0$ and $f_0$ systems specific choices for $V_0$ should be made, and that leads to the second approach which is the topic of the present work.

In approach 2,  a specific choice for the chiral invariant part of $V_0$ is made \cite{fjs_glsm,global} (in addition to modeling the axial anomaly, the $V_{SB}$ and the condensates).   The main conclusions are: (i) Again, light pseudoscalars have the tendency of   becoming quark-antiquark states;
(ii)Light scalars tend to become mainly four-quark states; and (iii)Predictions for various low-energy processes such as $\pi\pi$, $\pi K$, $\pi \eta$ scatterings, and decays such as $\eta'$ decays or semileptonic decays of $D_s$ can be made.  The rest of this article is devoted to a review of approach 2.

First, how is $V_0$ modeled?    Obviously, there are infinite number of terms that can be written down for $V_0$.   Up to dimension 4, there are twenty one SU(3)$_{\rm L}\times$ SU(3)$_{\rm R}$ invariant
terms in $V_0$ which can be made out of $M$
and $M'$:
\begin{eqnarray}
V_0 =&-&c_2 \, {\rm Tr} (MM^{\dagger}) +{\tilde c}_3 \, ({\rm det} M +
{\rm {\rm h.c.}}) +
 c_4^a \, {\rm Tr} (MM^{\dagger}MM^{\dagger})
 + c_4^b \, \left( {\rm Tr} (MM^\dagger)\right)^2
\nonumber \\
 &+& d_2 \,
{\rm Tr} (M^{\prime}M^{\prime\dagger})
     + d_3 \, ( {\rm det} M' + {\rm {\rm h.c.}})
      + d_4^a \, {\rm Tr} (M' M'^ \dagger M' M'^ \dagger)
     + d_4^b \, \left( {\rm Tr}
(M^{\prime}M^{{\prime}{\dagger}})\right)^2
\nonumber \\
     &+& e_2 \, ( {\rm Tr} (MM^{\prime\dagger})+
{\rm {\rm h.c.}})
 + e_3^a(\epsilon_{abc}\epsilon^{def}M^a_dM^b_eM'^c_f +
{\rm h.c.}) +
  e_3^b(\epsilon_{abc}\epsilon^{def}M^a_dM'^b_eM'^c_f +
{\rm h.c.})
\nonumber \\
     &+& e_4^a \, {\rm Tr} (MM^{\dagger}M^{\prime}M^{{\prime}{\dagger}})
      + e_4^b \, {\rm Tr} (MM^{{\prime}{\dagger}}M^{\prime}M^{\dagger})
     + e_4^c \, [{\rm
Tr} (MM^{\prime\dagger}MM^{\prime\dagger})+ {\rm h.c.}]
     + e_4^d \, [{\rm Tr} (M M^\dagger M M'^\dagger) +
{\rm h.c.}]
\nonumber \\
 &+& e_4^e \, [{\rm Tr} (M'M'^\dagger M'M^\dagger) +
{\rm h.c.}]
     + e_4^f \, {\rm Tr} (M M^\dagger) {\rm Tr} (M'M'^\dagger)
      + e_4^g \, {\rm Tr} (M M'^\dagger) {\rm Tr} (M'M^\dagger)
 \nonumber \\
      &+& e_4^h \, [( {\rm Tr} (M' M'^\dagger))^2 +
{\rm h.c.}]
+ e_4^i \, [ {\rm Tr}(M M^\dagger){\rm Tr}(M
M'^\dagger) + {\rm h.c.}]
      + e_4^j \, [ {\rm Tr} (M' M'^\dagger) {\rm Tr} (M'M^\dagger) + h.
c.].
\end{eqnarray}
\label{renormV0}
Notice that among these terms, those with the coefficients
$c_2$, $d_2$,  $c_4^a$, $c_4^b$, $d_4^a$, $d_4^b$, $e_3^a$, $e_4^a$, $e_4^b$, $e_4^f$, $e_4^g$ and
$e_4^h$ are $U(1)_A$ invariant.      As we go down the list of terms in $V_0$, we see that the number of quark and antiquark lines increases.  To work with such a high number of terms,  it seems reasonable to consider an approximation scheme in which $V_0$ is organized in terms of the number of quark and antiquark lines.    We define a parameter to keep track of this counting:
\begin{equation}
N=2 n + 4 n',
\end{equation}
where $n$ is the number of $M$ or $M^\dagger$ and $n'$ is the number of $M'$ or $M'^\dagger$ in a term.    In our first attempt we work in $N=8$ order which leads to:
\begin{eqnarray}
V_0^{(N\le 8)} =&-&c_2 \, {\rm Tr} (MM^{\dagger}) +{\tilde c}_3 \, ({\rm det} M +
{\rm {\rm h.c.}}) +
 c_4^a \, {\rm Tr} (MM^{\dagger}MM^{\dagger})
 + c_4^b \, \left( {\rm Tr} (MM^\dagger)\right)^2
\nonumber \\
 &+& d_2 \,
{\rm Tr} (M^{\prime}M^{\prime\dagger})
               + e_2 \, ( {\rm Tr} (MM^{\prime\dagger})+
{\rm {\rm h.c.}})
 + e_3^a(\epsilon_{abc}\epsilon^{def}M^a_dM^b_eM'^c_f +
{\rm h.c.}).
\end{eqnarray}
  The $c_4^b$ term has the structure of $\left({\rm Tr(\cdots)}\right)^2$ which can be shown to be inconsistent with OZI rule, therefore, we ignore this term at this level of investigation.  In addition, we mock up the $U(1)_{\rm A}$ anomaly exactly, which implies that the two terms ${\tilde c}_3$ and $e_2$ have to be combined in a nonlinear form in terms of natural logs in the following form \cite{U1_A}:
  \begin{equation}
c_3\left[\gamma_1 {\rm ln}\, \left(\frac{{\rm det}\,
M}{{\rm det}
M^{\dagger}}\right)
 +(1-\gamma_1)
{\rm ln} \,
\left(\frac{{\rm Tr}\,
(MM'^{\dagger})}{{\rm Tr}\, (M'M^{\dagger})}\right)\right]^2.
\label{altanom}
\end{equation}

Similarly, there are infinite number of terms in $V_{SB}$.  Up to dimension 4, terms linear in the flavor symmetry breaking matrix $A={\rm diag.}(A_1, A_2, A_3)$, are:
\begin{eqnarray}
V_{SB} =&+&k_1[ {\rm Tr} (AM)+ {\rm h.c.}]+k_2[ {\rm Tr}
(AM')+ {\rm h.c.}]
      +k_3[ {\rm Tr} (AMM^\dagger M) + {\rm h.c.}]+k_4[
{\rm Tr}
(AMM'^\dagger M')+ {\rm h.c.}]
\nonumber \\
      &+&k_5[ {\rm Tr} (AMM^\dagger M')+
{\rm h.c.}]+k_6[{\rm Tr}
(AMM'^\dagger M)+ {\rm h.c.}]
+ k_7[{\rm Tr} (AM'M'^\dagger M') + {\rm h.c.}]
\nonumber \\
&+&k_8[
{\rm Tr} (AM'M^\dagger
M)+
{\rm h.c.}]
+ k_9[ {\rm Tr} (AM'M'^\dagger M)+ {\rm
h.c.}]+k_{10}[{\rm Tr}
(AM'M^\dagger M')+ {\rm h.c.}]
\nonumber \\
   &+&k_{11}[{\rm Tr} (AM)+ {\rm h.c.}]{\rm
Tr}(MM^\dagger)
   + k_{12}[{\rm Tr} (AM)+ {\rm h.c.}]{\rm
Tr}(M'M'^\dagger)
  + k_{13}[{\rm Tr} (AM) {\rm Tr}(MM'^\dagger) +
{\rm h.c.}]
\nonumber \\
&+& k_{14}[{\rm Tr}
(AM)
{\rm Tr}(M'M^\dagger) + {\rm h.c.}]
+ k_{15}[{\rm Tr} (AM')+ {\rm h.c.}]{\rm
Tr}(MM^\dagger)
+ k_{16}[{\rm Tr} (AM')+ {\rm h.c.}]{\rm
Tr}(M'M'^\dagger)
\nonumber \\
   &+&k_{17}[{\rm Tr} (AM'){\rm Tr}(MM'^\dagger) +
{\rm h.c.}]+k_{18}[{\rm Tr}
(AM'){\rm Tr}(M'M^\dagger) + {\rm h.c.}]
+ k_{19} [A_a^b \epsilon_{bcd}\epsilon^{aef} M_e^c M_f^d +
{\rm h.c.}]
\nonumber \\
   &+&k_{20} [A_a^b \epsilon_{bcd} \epsilon^{aef} {M'}_e^c
{M'}_f^d + {\rm h.c.}]
+ k_{21} [A_a^b \epsilon_{bcd} \epsilon^{aef}M_e^c
{M'}_f^d + {\rm h.c.}].
\label{renormVsb}
\end{eqnarray}
In our first approximation, we consider only the first term in the right hand side of Eq. (\ref{renormVsb}) that is consistent with the quark mass term in the QCD Lagrangian.    Therefore,  at this level of approximation, the potential is:
\begin{eqnarray}
V =&-&c_2 \, {\rm Tr} (MM^{\dagger}) +
 c_4^a \, {\rm Tr} (MM^{\dagger}MM^{\dagger})
 + d_2 \,
{\rm Tr} (M^{\prime}M^{\prime\dagger})
 + e_3^a(\epsilon_{abc}\epsilon^{def}M^a_dM^b_eM'^c_f +
{\rm h.c.})
\nonumber \\
&+& c_3\left[\gamma_1 {\rm ln}\, \left(\frac{{\rm det}\,
M}{{\rm det}
M^{\dagger}}\right)
 +(1-\gamma_1)
{\rm ln} \,
\left(\frac{{\rm Tr}\,
(MM'^{\dagger})}{{\rm Tr}\, (M'M^{\dagger})}\right)\right]^2
+ k_1[ {\rm Tr} (AM) +{\rm h.c.}].
\label{V_N8}
\end{eqnarray}
We assume isospin limit which implies:
\begin{eqnarray}
A_1 &=& A_2 \ne A_3, \nonumber\\
\alpha_1 &=& \alpha_2 \ne \alpha_3, \nonumber\\
\beta_1 &=& \beta_2 \ne \beta_3,
\end{eqnarray}
where $\alpha$'s and $\beta$'s are the two- and four-quark condensates of fields $S$ and $S'$, respectively:
\begin{eqnarray}
\langle S_a^b \rangle &=& \delta_a^b \alpha_a,\nonumber \\
\langle {S'}_a^b \rangle &=& \delta_a^b \beta_a.
\end{eqnarray}
\section{Some Numerical Results}
At the level of the approximate potential of Eq. (\ref{V_N8}), altogether, there are 12 unknown parameters $c_2$, $c_4^a$, $d_2$, $e_3^a$, $c_3$, $\gamma_1$, $\alpha_1$, $\alpha_3$, $\beta_1$, $\beta_3$, $A_1$, $A_3$ that need to be determined using 12 inputs.   We take the following eight experimental inputs:
\begin{eqnarray}
 m[a_0(980)] &=& 984.7 \pm 1.2\, {\rm MeV},
\nonumber
\\ m[a_0(1450)] &=& 1474 \pm 19\, {\rm MeV},
\nonumber \\
 m[\pi(1300)] &=& 1300 \pm 100\, {\rm MeV},
\nonumber \\
 m_\pi &=& 137 \, {\rm MeV},
\nonumber \\
F_\pi &=& 131 \, {\rm MeV},
\nonumber \\
{A_3\over A_1} &=& 20\rightarrow 30, \nonumber \\
 \det(M_\eta^2) &=& \det(M_\eta^2)^{\rm exp.}, \nonumber \\
 {\rm Tr} (M_\eta^2) &=& {\rm Tr} (M_\eta^2)^{\rm exp.},
\label{inputs1}
\end{eqnarray}
together with four minimum conditions:
\begin{equation}
\left\langle\frac{\partial V}{\partial S_1^1}\right\rangle =
\left\langle\frac{\partial V}{\partial S_3^3}\right\rangle =
\left\langle\frac{\partial V}{\partial {S'}_1^1}\right\rangle =
\left\langle\frac{\partial V}{\partial {S'}_3^3}\right\rangle = 0.
\end{equation}
These inputs and minimum conditions allow a determination of the Lagrangian parameters and the detailed numerical analysis of this study in given in \cite{global}.  The main uncertainties are clearly on $m[\pi(1300)]$ as well as the ratio ${A_3\over A_1}$,  and therefore,  the results will have some variations that reflect these two uncertainties.    The prediction of the model for a  typical solution is given in the Tables \ref{phi_content_1215} and \ref{s_content_1215}.
\begin{table}[htbp]
\begin{center}
\begin{tabular}{c|c|c|c}
\hline \hline
State & \,\, ${\bar q} q$\% \, \, & \, ${\bar q}
{\bar q} q q$\%\, & \, $m$ (GeV)
\\
\hline
\hline
$\pi$      & 85  & 15 & \underline{0.137}\\
\hline
$\pi'$      & 15  & 85 & \underline{1.215}\\
\hline
$K$          & 86  & 14 & 0.515\\
\hline
$K'$          & 14  & 86 & 1.195\\
\hline
$\eta_1$   &  89   &  11   & 0.553 \\
\hline
$\eta_2$   &  78   &  22   & 0.982\\
\hline
$\eta_3$   &  32   &  68   &  1.225 \\
\hline
$\eta_4$   &  1   & 99     & 1.794\\
\hline
\hline
\end{tabular}
\end{center}
\caption[]{
Typical predicted properties of pseudoscalar
states:
${\bar q} q$ percentage (2nd
column), ${\bar q} {\bar q} q q$ (3rd
column) and masses (last column).  The two underlined masses in the last column are inputs and the rest of the numbers are predictions. The results are obtained for $m[\Pi(1300)]$ = 1.215 GeV.}
\label{phi_content_1215}
\end{table}
\begin{table}[htbp]
\begin{center}
\begin{tabular}{c|c|c|c}
\hline \hline
State &\,\, ${\bar q} q$\%\,\,& \, ${\bar q}
{\bar q} q q$\% \, & \, $m$ (GeV)
\\
\hline
\hline
$a$        &  24 &  76 & \underline{0.984}   \\
\hline
$a'$        &  76 &  24 & \underline{1.474}   \\
\hline
$\kappa$  &   8  &  92  & 1.067  \\
\hline
$\kappa'$  &   92  &  8  & 1.624 \\
\hline
$f_1$     &  40    &  60  & 0.742  \\
\hline
$f_2$     &  5   &    95  & 1.085 \\
\hline
$f_3$     &  63   &   37  & 1.493 \\
\hline
$f_4$     &  93   &   7   & 1.783 \\
\hline
\hline
\end{tabular}
\end{center}
\caption[]{ Typical predicted properties of scalar
states:
${\bar q} q$ percentage (2nd
column), ${\bar q} {\bar q} q q$ (3rd
column) and masses (last column). The two underlined masses in the last column are inputs and the rest of the numbers are predictions. The results are obtained for $m[\Pi(1300)]$ = 1.215 GeV.}
\label{s_content_1215}
\end{table}
We see in Table \ref{phi_content_1215} that the underlying mixing among two- and four-quark pseudscalars does not change the well-established picture of the light pseudoscalars as dominantly being quark-antiquark states.  The situation is of course reversed for the heavy pseudoscalars.    The present level of this investigation predicts four etas:   The two below 1 GeV ($\eta_1$ and $\eta_2$) are clearly close to the physical states $\eta(547)$ and $\eta(958)$.  The two heavier etas above 1 GeV ($\eta_3$ and $\eta_4$) may be identified with two of the several physical eta states above 1 GeV.    In the work of \cite{global} various scenarios for this identification is studied in detail and the closest agreement corresponds to identifying $\eta_3$ with $\eta(1294)$ and $\eta_4$ with $\eta(1617)$.

\begin{figure*}[ht]
\centering
\includegraphics[width=100mm]{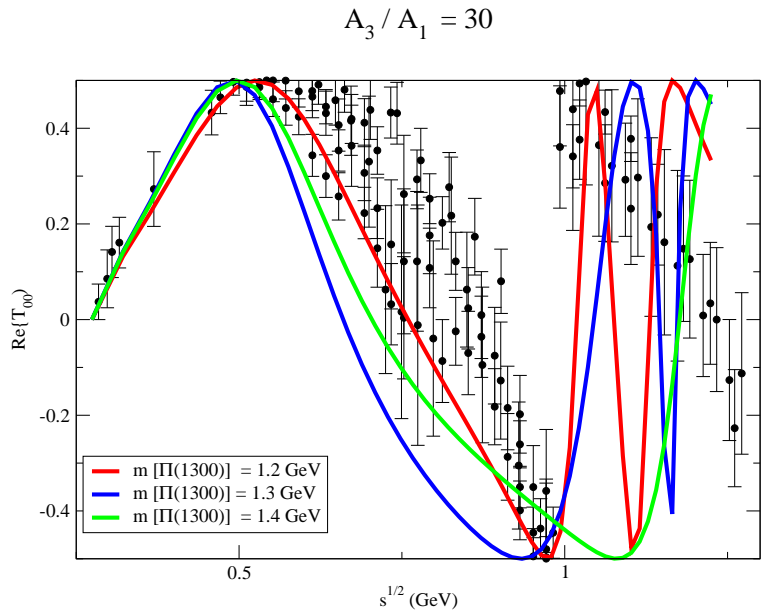}
\caption{Prediction of the present mixing approximation for the real part of the I=J=0 $\pi\pi$ scattering amplitude unitarized by K-matrix method.} \label{F_ReT00}
\end{figure*}

The results for scalars are given in Table \ref{s_content_1215} and overall are the inverse of those of pseudoscalars:    Light scalars below 1 GeV tend to gain a large four-quark component (and vice versa for heavy scalars above 1 GeV).
The predicted ``Lagrangian masses'' for the scalars are higher than their expected values. However, they receive corrections when probed in appropriate unitarized scattering amplitudes.   For example, the Lagrangian squared-masses for the four isosinglet scalar states  ($f_1$, $f_2$, $f_3$ and $f_4$) appear as poles in the $\pi\pi$ scattering amplitude and differ from the poles in the K-matrix unitarized $\pi\pi$ scattering amplitude.    When physical masses and decay widths are extracted from the poles in the unitarized amplitudes, it is found that these masses and decay widths are considerably closer to their expected values.  Table \ref{T_poles} presents the four physical masses and decay widths found from the poles of the K-matrix unitarized $\pi\pi$ scattering amplitude in ref. \cite{fjss11_poles}.  The first physical mass and width is clearly consistent with the property of sigma meson; the second mass and width is qualitatively close to those of $f_0(980)$ and the remaining two masses and widths may be compared with the masses  and widths of two of the several isosinglet states above 1 GeV.  However,  at this level of investigation we do not expect the properties of the two physical states
above 1 GeV to be accurate due to the neglect (for simplicity) of several important factors, including $K{\bar K}$ threshold and mixing with glueballs.    The predicted K-matrix unitarized $\pi\pi$ scattering amplitude is given in Fig. \ref{F_ReT00} and we see a reasonable agreement up to about 1 GeV.
\begin{table}[htbp]
\begin{center}
\begin{tabular}{c||c|c||c|c}
\hline \hline
Pole & Mass (MeV) & Width (MeV) & Mass (MeV) &
Width (MeV)
\\ \hline
1 & 483 & 455 & 477 & 504
\\ \hline
2 & 1012& 154 & 1037 & 84
\\ \hline
3 & 1082 & 35 & 1127 & 64
\\ \hline
4 & 1663 & 2.1 & 1735 & 3.5
\\ \hline
\hline
\end{tabular}
\end{center}
\caption[]{The physical mass and decay width of
the isosinglet scalar states, with $m[\Pi(1300)]
=1.215$ GeV and with $A_3/A_1$ = 20 (the
first two columns) and with $A_3/A_1$ = 30
(the last two columns).}
\label{T_poles}
\end{table}
\begin{figure*}[ht]
\centering
\includegraphics[width=100mm]{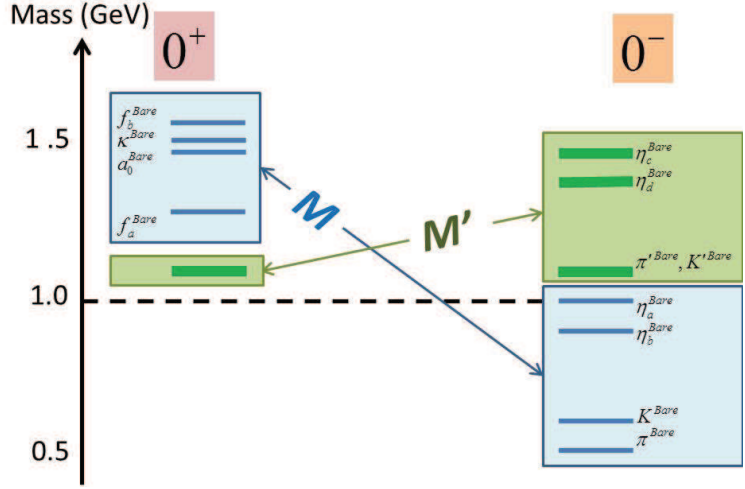}
\caption{The ``bare'' mass spectrum obtained in the present order of approximation ($N\le 8$) of the $MM'$ mixing model (with $m[\Pi(1300)]=1.215$ GeV).  The two-quark chiral nonet $M$ is shown in blue and the four-quark chiral nonet $M'$ is shown in green. }
\label{F_bare_mass_spectrum}
\end{figure*}
\section{Summary and Conclusion}
This review presented a mixing picture for scalar and pseudoscalar mesons below and above 1 GeV.   We saw that the generalized linear sigma model provides an appropriate framework for studying this mixing and allows an estimate of the substructure of scalar and pseudoscalar mesons.  Allowing a mixing between a two- and a four-quark chiral nonets,  the scalars below 1 GeV come out close to four-quark states (whereas those above 1 GeV come out closer to two-quark states) and vice versa for pseudoscalar mesons.   A summary of this chiral mixing model in presented in Fig. \ref{F_bare_mass_spectrum} in which it is shown how at the present level of approximation two- and four-quark nonets are formed out of ``bare'' states.  Further extensions of this model is desirable:  Higher order $N$ values are expected to improve the estimates;  inclusion of glueballs will allow a more reliable determination of the isosinglet states above 1 GeV (both scalars and pseudoscalars); and the effect of K-matrix unitarization on $\pi K$ channel is expected to reveal the  properties of kappa meson.
\begin{acknowledgments}
The author wishes to thank the organizers of DPF 2011 for a very productive conference.    This work is based on an ongoing collaboration with R. Jora, J. Schechter and M.N. Shahid, and the author thanks collaborators for many helpful discussions.  This work has been partially supported by the NSF Grant 0854863 and by a 2011 grant from
the Office of the Provost, SUNYIT.
\end{acknowledgments}

\end{document}